\documentclass[pdflatex,sn-basic]{sn-jnl}

\usepackage{graphicx}%
\usepackage{amsmath,amssymb,amsfonts}%
\usepackage{booktabs}%
\usepackage{tabularx}%
\usepackage{multirow}%
\usepackage{longtable}%
\usepackage{float}%
\usepackage{placeins}%
\usepackage{xcolor}%
\usepackage{setspace}

\hypersetup{bookmarksdepth=3,hypertexnames=false}
\graphicspath{{./}}

\newcommand{\NetCountries}{114}

\newcommand{\NetPairs}{272}
\newcommand{\NetEdges}{84}
\newcommand{\NetSynergy}{40}
\newcommand{\NetTradeoff}{44}
\newcommand{\NetDensity}{0.31}

\newcommand{\NetMaxExcluded}{20}
\newcommand{\NetMeanExcluded}{2.5}
\newcommand{\NetLagThreeOverlap}{82}
\newcommand{\NetLagThreeSign}{81}

\newcommand{\NetCCESpearman}{0.81}
\newcommand{\NetCCECore}{72}

\newcommand{\NetBYEdges}{59}
\newcommand{\NetBYShare}{70}
\newcommand{\NetBYSynergy}{25}
\newcommand{\NetBYTradeoff}{34}
\newcommand{\NetBYDensity}{0.22}
\newcommand{\LPPairs}{31}
\newcommand{\LPSupported}{2}
\newcommand{\LPSuggestive}{3}

\newcommand{\LPHorizonStar}{5}
\newcommand{\CipsGoals}{17}
\newcommand{\CipsLevelUnitRoot}{11}
\newcommand{\CipsDiffStationary}{17}
\newcommand{\CIPScv}{-2.18}
\newcommand{\NetDegSpearKOne}{0.03}
\newcommand{\NetDegSpearKThree}{0.18}
\newcommand{\NetWtSpearKOne}{0.19}
\newcommand{\NetWtSpearKThree}{0.21}
\newcommand{\NetBootB}{500}
\newcommand{\PDrvNoPoverty}{0.98}
\newcommand{\PDrvPartnerships}{0.82}
\newcommand{\PDrvCities}{0.76}
\newcommand{\PDrvEducation}{0.54}

\newcommand{\PDrvPeace}{0.02}
\newcommand{\PDrvConsumption}{0.35}
\newcommand{\NoPovertyCIlow}{+0.06}
\newcommand{\NoPovertyCIhigh}{+0.47}
\newcommand{\PeaceCIlow}{-0.73}
\newcommand{\PeaceCIhigh}{-0.07}
\newcommand{\NetAICShareKOne}{66}
\newcommand{\NetBICShareKOne}{83}
\newcommand{\SubRetainedPreSDG}{57}

\newcommand{\SubRetainedPreCovid}{62}

\newcommand{\BndRetainedShare}{87}
\newcommand{\BndSignAgree}{93}
\newcommand{\BndTradeShareChange}{+0.02}

\newcommand{\OvlIndicators}{126}

\newcommand{\OvlRho}{0.99}

\newcommand{\SelPopShare}{89}
\newcommand{\SelMissingIncl}{5.1}
\newcommand{\SelMissingExcl}{15.4}
\newcommand{\SelMissingD}{-0.88}

\newcommand{\HetMedianConsensus}{54}
\newcommand{\WdiBeta}{-0.044}
\newcommand{\WdiCIlow}{-0.066}
\newcommand{\WdiCIhigh}{-0.021}

\newcommand{\WdiPerTenPP}{-0.44}
\newcommand{\WdiCountries}{183}
\newcommand{\WdiObs}{2861}
\newcommand{\PovRevBeta}{+0.34}
\newcommand{\PovRevP}{0.008}
\newcommand{\SanRevP}{0.75}
\makeatletter
\newcommand{\PrismSupplementLabel}[2]{\expandafter\gdef\csname r@#1\endcsname{{#2}{}}}
\PrismSupplementLabel{tab:coverage}{S1}
\PrismSupplementLabel{tab:cips}{S2}
\PrismSupplementLabel{tab:edges}{S3}
\PrismSupplementLabel{tab:subperiod}{S4}
\PrismSupplementLabel{tab:selection}{S5}
\PrismSupplementLabel{tab:lp}{S6}
\PrismSupplementLabel{tab:placebo}{S7}
\PrismSupplementLabel{tab:strata}{S8}
\PrismSupplementLabel{tab:wdi}{S9}
\makeatother

\newcommand{\SDGs}{Sustainable Development Goals}

\begin{document}

\title[Adaptive Sustainability Portfolios]{Beyond Sustainable Development Goal (SDG) Accelerators: Directed Evidence for Adaptive Sustainability Portfolios}

\author[1]{\fnm{Md Muhtasim Munif} \sur{Fahim}}\email{fahim.stat.ru@gmail.com}
\author[1]{\fnm{Md Jahid Hasan} \sur{Imran}}\email{mjh.imran60@gmail.com}
\author[1]{\fnm{Md. Naim} \sur{Molla}}\email{naim.molla.stats@gmail.com}
\author[1]{\fnm{Luknath} \sur{Debnath}}\email{luknath57@gmail.com}
\author[1]{\fnm{Tonmoy} \sur{Shil}}\email{tonmoy.stat.ru@gmail.com}
\author[1]{\fnm{Ehsanul Bashar} \sur{Pranto}}\email{s2110924113@ru.ac.bd}
\author[1]{\fnm{Md Mostafizur Rahman} \sur{Likhon}}\email{mrlikhon813@gmail.com}
\author[1]{\fnm{Md Shafin Sanyan} \sur{Saad}}\email{saadshafin0@gmail.com}
\author*[1]{\fnm{Md. Rezaul} \sur{Karim}}\email{mrkarim@ru.ac.bd}

\affil[1]{\orgdiv{Data Science Research Lab, Department of Statistics}, \orgname{University of Rajshahi}, \orgaddress{\city{Rajshahi}, \postcode{6205}, \country{Bangladesh}}}

\abstract{Governments and development partners often seek Sustainable Development Goal (SDG) ``accelerators'' whose progress can propagate gains across human well-being, institutions, resources, and environmental priorities. Such claims require evidence on direction, timing, robustness, and trade-offs, not only network connectivity. We map directed SDG interdependence using a balanced panel of \NetCountries\ countries observed annually from 2000 to 2024. Because the goal series are persistent, trending, and cross-sectionally dependent, we combine Dumitrescu--Hurlin panel Granger non-causality tests on first-differenced series with false-discovery control across \NetPairs\ directed goal pairs, and panel local projections with Driscoll--Kraay standard errors for \LPPairs\ theory-derived indicator linkages. We find \NetEdges\ directed linkages after correction: \NetSynergy\ synergies and \NetTradeoff\ trade-offs. This dense structure does not support a universal accelerator. Driver--receiver rankings are fragile across lag orders, centrality metrics, bootstrap resampling, and alternative dependence controls; only peace and strong institutions as a net receiver and poverty reduction as an effect-size-weighted driver are distinguishable in the bootstrap screen. Two local-projection pathways are supported, both leading to lower child mortality, and the education--child-health association is corroborated in independent World Development Indicators data across \WdiCountries\ countries. The policy implication is not to abandon SDG integration, but to replace fixed global accelerator rankings with adaptive portfolios built around supported, time-lagged linkages and monitored constituent indicators.}


\maketitle

\section{Introduction}

The 2030 Agenda asks governments to make progress on a broad set of social,
economic, environmental, and institutional objectives at the same time.
Implementation, however, occurs under binding fiscal, political, and
administrative constraints. This creates a recurring decision problem: should
scarce resources be concentrated on a small number of goals expected to
generate wider gains, or distributed across a portfolio whose components are
adjusted as evidence changes? The first option is attractive because it turns a
complex coordination problem into a ranking. The ranking, however, carries a
demanding evidential burden. A goal promoted as a global accelerator, one
whose progress is expected to pull other goals along, must do more than
correlate with them: it must \emph{drive} them, in a direction and on a
timescale that policy can act upon.

The SDG-interlinkage literature has established beyond reasonable dispute that
development outcomes are connected. Expert frameworks map how targets may
reinforce or constrain one another \citep{nilsson2016map,nilsson2018mapping};
correlation and network studies identify recurring empirical associations
\citep{pradhan2017systematic,swain2021modeling}; and recent work shows that
apparent priorities change with income, development stage, geography, and
time \citep{lusseau2019income,wu2022decoupling,huan2026spatiotemporal}.
Together, these findings give integrated policy a firm footing. What they do
not reveal, on their own, is the \emph{directed} structure of the system:
which goals propagate progress to others, which mostly absorb it, and how
long the transmission takes. Connectivity (goals move together),
directionality (which goal leads), and dynamics (over what horizon) are
distinct empirical claims, and an accelerator argument rests on the last two.

Recovering directionality from a global panel demands estimators matched to the
data. The SDG composites are highly persistent and trending, countries
co-move through shared global shocks, and most panels are wide but only
moderately long. These features defeat the short-panel
generalized-method-of-moments estimators often applied to such data, whose
instrument counts proliferate when the time dimension is long
\citep{roodman2009note}. They also make causal inference on series levels
vulnerable to common-trend spurious association
\citep{granger1974spurious}. Testing many directed pairs raises a further
problem: the more hypotheses are tested, the more will reach significance by
chance alone. With seventeen goals there are \NetPairs\ ordered cross-goal
relationships, so an uncorrected significance threshold can manufacture an
apparently meaningful network out of false positives. A credible directed map
therefore requires methods built for wide, trending, cross-sectionally
dependent panels, inference on stationary-transformed series, and explicit
false-discovery control.

This paper maps the directed interdependence structure of all seventeen
\SDGs\ using a balanced panel of \NetCountries\ countries observed annually from 2000 to
2024. We use two estimators matched to this regime. A Dumitrescu--Hurlin panel
Granger non-causality test \citep{dumitrescu2012}, run on first-differenced
series with false-discovery control, recovers the directed network of
goal-to-goal influence and ranks goals by net influence. Panel local
projections \citep{jorda2005} with Driscoll--Kraay standard errors then measure
the dynamic magnitude of \LPPairs\ theory-derived indicator linkages drawn
from recent synthesis studies
\citep{laumann2022,swain2021modeling,bennich2023recurring}. The
analysis applies a fixed set of estimators, lag orders, stationarity
treatments, and false-discovery families, declared before the confirmatory
runs. Each linkage is then graded in an explicit claim vocabulary as
supported, suggestive, or unsupported, and enters the abstract or the policy
discussion only if it survives these rules.

The resulting picture is both structured and disciplined. Of \NetPairs\ directed goal
pairs, \NetEdges\ linkages survive false-discovery control (\NetSynergy\ synergies and \NetTradeoff\
trade-offs; network density \NetDensity). The system is densely and directionally
connected, and because synergies and trade-offs occur at comparable strength,
no single goal behaves as a universal accelerator. We then ask a question this
literature rarely puts to its own outputs: is the goal-level driver--receiver
hierarchy itself a robust object? It is not. Rankings of net influence
correlate only weakly across lag orders (Spearman \NetDegSpearKOne--\NetWtSpearKThree)
and across centrality metrics. Under a country bootstrap, only two goals hold
roles whose intervals exclude zero: Peace and Institutions as the system's
clearest net receiver, and No Poverty as the most probable driver under the
effect-size-weighted metric. Partnerships and sustainable cities lean toward
driver status under every metric without reaching significance. What does survive
every check is a small set of specific dynamic linkages: of the \LPPairs\
prespecified indicator pairs, \LPSupported\ are supported and \LPSuggestive\
suggestive, with effects accruing over four to five years. The strongest are
improvements in sanitation and in poverty reduction preceding lower child
mortality; both survive false-discovery control and hold in poorer and richer
country strata alike. The association between education and child health also
appears in independent World Development Indicators data.

These findings make three contributions. Methodologically, they show how to
recover a directed, dynamically resolved SDG interaction map from a macro panel
using estimators matched to the data, and how to stress-test the goal-level
rankings such maps produce. The demonstration that driver lists are fragile to
lag order and centrality metric applies with equal force to earlier
interlinkage hierarchies built on unweighted degree, that is, on simple
counts of a goal's significant links irrespective of their strength.
Substantively, the findings replace the search for a single master goal, and
for a master ranking, with a small set of empirically supported, time-lagged
linkages embedded in a dense but heterogeneous web. For policy, we read these
results as grounds to reframe the prioritization problem around those
specific linkages and their multi-year horizons: anchor investment cases on
supported pathways rather than on goal-level league tables, monitor the
constituent indicators expected to move, and allow several years for
transmission. This is consistent with recent
multiobjective portfolio work \citep{yang2025portfolios} while remaining
explicit about what an observational global panel can and cannot establish.
\section{From Interlinkage Maps to Directed, Dynamic Evidence}

\subsection{Four evidentiary levels}

The term ``interlinkage'' is used for several objects that should not be
treated as interchangeable. At the first level, two outcomes move together.
Pairwise correlation, expert coding, and descriptive network edges can be
valuable for identifying coordination problems, but they do not establish
temporal ordering. At the second level, a lagged change in one measure improves
prediction of a subsequent change in another after controlling for persistence
and common components, and a directed network can be assembled from these tests.
Panel Granger non-causality and local-projection models operate at this level,
the first establishing the existence and direction of a linkage and the second
its dynamic magnitude. At the third
level, the proposed relationship appears in independent constituent-indicator
data. This is a mechanism-consistency test: it asks
whether the aggregate pattern has a recognizable empirical counterpart. At the
fourth level, a feasible policy intervention changes the outcome through the
claimed mechanism. That level requires a design capable of identifying an
intervention effect and lies outside this study.

Confusing these levels encourages overstatement. A central goal in a
correlation network may not conditionally predict other goals. A conditional
predictor among composite scores may fail when represented by enrollment,
mortality, poverty, or expenditure indicators. A directed Granger edge among composite
scores may weaken or reverse when the goals are represented by their
constituent indicators. We therefore use ``Granger-causal linkage'' for the
network edges, ``dynamic response'' for the local-projection estimates, and
``mechanism consistency'' for the independent-indicator triangulation. None is
described as a structural intervention effect.

\subsection{Why universal rankings are fragile}

Three features of SDG data make universal rankings especially vulnerable.
First, goal scores are composites. The indicators used to construct them differ
in scale, variability, missingness, and policy proximity. Aggregation can
increase signal by averaging noise, but it can also create relationships that
do not correspond to any single mechanism. A change in an education score, for
example, need not mean the same combination of enrollment, completion, or
learning changes in every country. A relationship among goal scores is thus a
relationship among measurement systems as well as among substantive outcomes.

Second, global panels combine heterogeneous trajectories. Income groups,
regions, conflict exposure, demographic transitions, and statistical capacity
all shape observed change. Prior studies find that network structures differ
across income groups \citep{lusseau2019income}, development stages
\citep{wu2022decoupling}, and spatial or temporal samples
\citep{huan2026spatiotemporal}. These findings are not anomalies around a fixed
global network. They imply that the global average can conceal distinct local
relationships. A universal accelerator is therefore a much stronger claim
than the existence of an average association.

Third, the network is selected through many simultaneous tests. With \NetPairs\
ordered cross-goal pairs, roughly fourteen false positives are expected on
average at a five-percent threshold under a simple global null. The dependence
among tests does not make this arithmetic an exact prediction, but it makes
uncorrected edge counts poor evidence for a policy ranking. False-discovery
control is not an optional robustness exercise after a network has been
interpreted. It is part of defining which edges are eligible for
interpretation, and it is why the directed network we report is substantially
sparser than the raw test count would suggest.

\subsection{Position in the literature}

Early SDG frameworks emphasized systematic mapping before integrated policy
decisions \citep{nilsson2016map}. Subsequent studies used global indicators to
identify recurring associations \citep{pradhan2017systematic}, regional
communities \citep{swain2021modeling}, and context-dependent centrality
\citep{lusseau2019income}. Review work now documents both the practical value
and the methodological diversity of these approaches
\citep{bennich2023recurring,issa2024network,khot2026gaps}. More recent
directional analyses infer dense temporal structures at goal and indicator
levels \citep{bazah2026pillars,ding2026global}. Multi-criteria studies, in
parallel, argue that prioritization must combine gaps, interactions, and
decision context rather than rely on centrality alone
\citep{allen2019prioritising,asadikia2024navigating}.

The present contribution is to recover the \emph{directed and dynamic}
structure of the goal system with estimators matched to a macro panel, under
explicit false-discovery control. Rather than a correlation or expert-coded map,
we report which goals Granger-drive others across all seventeen goals, separate
net drivers from net receivers, and quantify how long each supported linkage
takes to act. This responds to two gaps identified in recent reviews: the
dominance of undirected, goal-level evidence and the weak resolution of the
timescale over which interactions operate \citep{issa2024network,khot2026gaps}.
The reporting standard remains asymmetric. A linkage that survives diagnostics,
multiplicity, and, where testable, independent triangulation can inform
priorities. An edge that survives none of these cannot, however visually
prominent it appears in an uncorrected graph.
\section{Data and Methods}

\subsection{Data}

The analysis uses the Sustainable Development Report 2025 backdated SDG Index
\citep{sachs2025sdr}, which provides annual normalized scores (0--100, higher is
better) for all seventeen \SDGs\ and 102 constituent indicators. In the
backdated series the current indicator set, normalization bounds, and
aggregation rules \citep{lafortune2018sdg} are applied retrospectively to the
whole 2000--2024 window, so year-to-year changes reflect movement in the
underlying indicators rather than methodology revisions. We construct a
balanced panel of the \NetCountries\ countries with complete observations on all seventeen
goal composites in every year from 2000 through 2024, yielding 2{,}850
country-year rows with no missing values. Country-year uniqueness, workbook
integrity, source hashes, units, and variable-level coverage are verified before
estimation. Working with all seventeen goals rather than a reduced subset is
deliberate: the central object of study is the directed interdependence structure
of the goal system as a whole.

Indicator-level dynamics are estimated for a prespecified set of \LPPairs\
directed indicator pairs drawn from the mechanisms claimed in three recent
synthesis studies of SDG interactions
\citep{laumann2022,swain2021modeling,bennich2023recurring}.
Each pair maps one published source--target linkage onto a specific pair of
normalized SDR indicators (for example, sanitation access and child stunting).
For each pair we build a separate balanced panel of countries with complete
observations on both indicators in all twenty-five years. The pairs and their
literature provenance are frozen before estimation and listed in the supplement.
Supplementary Table~\ref*{tab:coverage} summarizes data sources and coverage.

\subsection{Choice of estimator for a macro panel}

The panel is wide and moderately long: $N \approx \NetCountries$ countries observed over
$T = 25$ years. This regime is poorly suited to difference and system
generalized method of moments estimators, which are designed for short panels
and whose instrument count grows with $T^{2}$; at $T=25$ the instrument matrix
proliferates and collapses even under standard remedies \citep{roodman2009note}.
The SDG composites are also highly persistent and trending, so causal inference
on their levels is contaminated by common-trend spurious association, and the
countries are cross-sectionally dependent through shared global shocks. We
therefore adopt two estimators matched to a wide, moderately long, trending,
cross-sectionally dependent panel: a heterogeneous panel Granger
non-causality test for the system-level interaction network, and panel local
projections for the dynamic magnitude of prespecified linkages. All inference is
conducted on first-differenced (year-over-year change) series to remove the
common trends that would otherwise generate spurious causality
\citep{granger1974spurious}. A Pesaran (2007) CIPS panel unit-root test
\citep{pesaran2007} supports this treatment. Computed from cross-sectionally
augmented Dickey--Fuller regressions with an intercept, no trend, and one
augmentation lag, it does not reject the unit-root null at 5\% for
\CipsLevelUnitRoot\ of \CipsGoals\ goal composites in levels, whereas all
\CipsDiffStationary\ are stationary in first differences (Supplementary
Table~\ref*{tab:cips}). The level evidence is thus mixed across composites, so
we use the test diagnostically rather than as a series-by-series
classification, and difference all seventeen composites uniformly as a
conservative, predeclared transformation that guarantees stationary inputs
and comparability across pairs.

\subsection{Directed interaction network: panel Granger non-causality}

We estimate the directed network of goal-to-goal influence with the
Dumitrescu--Hurlin panel test \citep{dumitrescu2012} of Granger
non-causality \citep{granger1969}, which permits fully heterogeneous
dynamics across countries and is valid for moderate-to-large $T$. For each ordered pair of goals $(x \rightarrow y)$ and
each country $i$ we estimate, in first differences,
\begin{equation}
\Delta y_{it}=\alpha_i+\sum_{k=1}^{K}\gamma_{ik}\,\Delta y_{i,t-k}
+\sum_{k=1}^{K}\beta_{ik}\,\Delta x_{i,t-k}+\varepsilon_{it},
\end{equation}
with lag order $K=2$, and form the individual Wald statistic $W_i$ for
$H_0:\beta_{i1}=\dots=\beta_{iK}=0$. The panel statistic is the cross-country
average $\bar{W}=N^{-1}\sum_i W_i$, standardized to the finite-sample
$\tilde{Z}$ of \citet{dumitrescu2012}, which is asymptotically standard normal
under the non-causality null. Because the test rejects only for large positive
$\tilde{Z}$ (that is, large $\bar{W}$), we use one-sided upper-tail $p$-values.
Country regressions with near-singular or degenerate designs (near-constant
differenced series, condition number above $10^{6}$, or non-finite Wald
statistics) are excluded so that a few numerically unstable units cannot
dominate $\bar{W}$. Across the \NetPairs\ directed goal pairs, Benjamini--Hochberg
false-discovery-rate control \citep{benjamini1995controlling} is applied to the
$p$-values, and a directed edge is retained at $q<0.05$. The sign of each
retained edge is the cross-country \emph{median} of $\sum_k \beta_{ik}$, robust
to outlying countries: positive denotes a synergy (co-improvement) and negative
a trade-off on the normalized scale. From the retained edges we compute each
goal's out-degree, in-degree, and net influence (out-degree minus in-degree),
which distinguishes net drivers from net receivers. Because the
Dumitrescu--Hurlin alternative is causality for \emph{some} countries, we also
report each edge's cross-country sign consensus (Supplementary Table~\ref*{tab:edges}): the
median consensus is \HetMedianConsensus\%, so the sign label describes the
median country rather than a universal direction. The numerical guards exclude
few units in practice (mean \NetMeanExcluded, maximum \NetMaxExcluded\ of
\NetCountries\ countries per pair).

The lag order $K=2$ was fixed in the frozen protocol to balance annual-data
dynamics against parameter cost. It is a prespecified choice, not an
information-criterion winner: per-country information criteria favor shorter
lags (AIC selects $K=1$ in \NetAICShareKOne\% and BIC in \NetBICShareKOne\% of
unit regressions), so we report lag sensitivity in full. Edge
\emph{membership} is moderately stable across longer lags
(\NetLagThreeOverlap\% of primary edges remain significant at $K=3$, with
\NetLagThreeSign\% sign agreement) but only about half survive at $K=1$, and
goal-level \emph{rankings} are not lag-stable: the Spearman correlation of net
influence with the primary network is \NetDegSpearKOne\ at $K=1$ and
\NetDegSpearKThree\ at $K=3$ under the degree metric, and remains weak
(\NetWtSpearKOne--\NetWtSpearKThree) under an effect-size-weighted metric. We
therefore quantify the uncertainty of every goal's role directly, recomputing
the weighted net influence in \NetBootB\ country-bootstrap resamples
(Table~\ref{tab:centrality}, Figure~\ref{fig:netinf}). Because countries co-move
through shared global shocks, we further verify that the network is not an
artifact of cross-sectional dependence. We re-estimate it after removing the
common shock by year-demeaning the differenced series (equivalent to time fixed
effects) and, more generally, after augmenting each country regression with
contemporaneous cross-sectional averages of the differenced series, the common
correlated effects estimator of \citet{pesaran2006cce}, which permits
country-specific factor loadings. A Pesaran CD test \citep{pesaran2015cd}
quantifies the dependence before and after each control, and we report the
stability of the driver--receiver ranking (Figure~\ref{fig:robust}B).

Four further reviewer-motivated checks complete the design. First, the SDR
codebook assigns each of its \OvlIndicators\ indicators to exactly one goal,
so no edge can be a mechanical artifact of shared indicators; the single
near-duplicate pair (marine and terrestrial protected areas, Goals 14--15)
changes one edge and leaves the net-influence ranking essentially unchanged
($\rho=\OvlRho$). Second, because the 0--100 scores are bounded, we re-estimate
the full network on logit-transformed scores: \BndRetainedShare\% of primary
edges are retained with \BndSignAgree\% sign agreement and an essentially
unchanged trade-off share (\BndTradeShareChange), and trade-off edges show no
excess concentration on near-ceiling goals. Third, we re-estimate the network
in sub-periods around the SDG adoption and COVID breaks
(Supplementary Table~\ref*{tab:subperiod}). Fourth, we characterize the balanced panel's
selection: exclusion is driven by data completeness (missing-data share
\SelMissingExcl\% versus \SelMissingIncl\%, Cohen's $d=\SelMissingD$) rather
than income, region, or governance composition, and the panel covers
\SelPopShare\% of 2024 world population (Supplementary Table~\ref*{tab:selection}).

\subsection{Dynamic magnitudes: panel local projections}

For the \LPPairs\ prespecified indicator linkages we estimate the dynamic
response of each target to a change in its source using panel local projections
\citep{jorda2005}. For horizons $h=0,1,\dots,10$ years we fit
\begin{equation}
y_{i,t+h}-y_{i,t-1}=\alpha_i^{h}+\lambda_t^{h}
+\beta_h\,\Delta s_{it}
+\sum_{l=1}^{2}\phi_l^{h}\,\Delta y_{i,t-l}
+\sum_{l=1}^{2}\psi_l^{h}\,\Delta s_{i,t-l}+e_{i,t+h},
\end{equation}
with country effects $\alpha_i^{h}$ and year effects $\lambda_t^{h}$. The
coefficient $\beta_h$ is the cumulative response of the target (in SDR score
points) to a one-point improvement in the source $h$ years earlier; the sequence
$\{\beta_h\}$ is the impulse-response function. Standard errors are
Driscoll--Kraay kernel estimates (Bartlett kernel with bandwidth
$\max(h,1)$), robust to heteroskedasticity, serial correlation, and
cross-sectional dependence across countries \citep{driscoll1998}. Local projections impose no instrument structure and so
avoid the proliferation that disables GMM in this panel. The confirmatory
estimand is the cumulative response at the prespecified \LPHorizonStar-year
horizon; its $p$-value enters a Benjamini--Hochberg family across the estimable
pairs, so the dynamic linkages are held to the same false-discovery standard as
the network. For pathways that pass this screen we add three diagnostics:
a lead placebo and a current-versus-lead horse race that probe timing, a
reverse-direction projection that probes feedback (Supplementary Table~\ref*{tab:placebo}),
and income-stratified re-estimation that probes pooling
(Supplementary Table~\ref*{tab:strata}, Supplementary Figure~S2).

\subsection{Claim classification}

We separate two evidentiary questions and report both. For the network, a
directed linkage is \emph{retained} when its Dumitrescu--Hurlin test survives
false-discovery control ($q<0.05$); retained edges are described by sign and
strength but are not interpreted as intervention effects. For the dynamic
linkages, an impulse response is \emph{supported} when its Benjamini--Hochberg
value across the estimable pairs at the \LPHorizonStar-year horizon is below
0.05 in the expected direction, and \emph{suggestive} when its raw $p$-value at
that horizon is below 0.05 but the false-discovery threshold is not met;
responses meeting neither rule are \emph{unsupported}. Pairs whose panels are
too small or numerically degenerate to estimate at the confirmatory horizon are
\emph{not interpretable}. Coefficient rows and companion
diagnostics are keyed by model identifier; all figures, tables, and reported
statistics read from this registry.

\subsection{Independent triangulation}

To check that the strongest indicator linkage is not an artifact of the SDR
construction, we re-estimate the education--health relationship in independent
data, combining World Development Indicators \citep{worldbank2026wdidataset} for
secondary enrollment and under-five mortality with an annual-change
specification,
\begin{equation}
\Delta z_{it}=\alpha+\theta\,\Delta x_{i,t-1}+\phi\,\Delta z_{i,t-1}
+\lambda_t+u_{it},
\end{equation}
with year effects $\lambda_t$ and standard errors clustered by country; the
first-difference form removes country-level effects, and a variant adds
country dummies (country-specific trends in changes). The estimation sample is
the full SDR country universe with available WDI data (\WdiCountries\
countries, \WdiObs\ country-years), deliberately broader than the balanced
network panel: the point of triangulation is independent data with its own
coverage. Series codes, controls (health expenditure, education expenditure,
GDP per capita, fertility), natural-unit effect sizes, and a balanced-panel
sensitivity subsample are reported in Supplementary Table~\ref*{tab:wdi}. This provides an
out-of-source check on the education-to-child-mortality pathway that also
appears in both the network and the local-projection results.
\section{Results}

\subsection{The SDG system is a dense, directional web}

\begin{figure}[htbp]
\centering
\includegraphics[width=\textwidth]{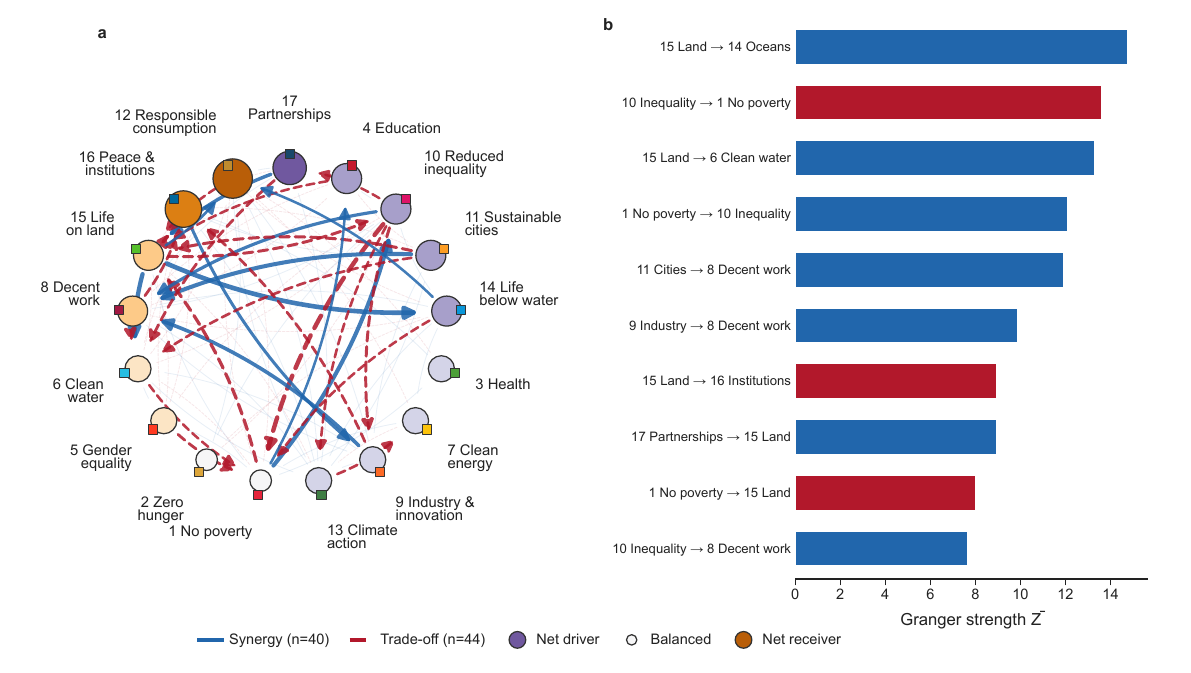}
\caption{Directed network of Granger-predictive SDG interlinkages, 2000--2024.
(\textbf{a})~Nodes are the seventeen goals, ordered clockwise from the
strongest net driver (top) to the strongest net receiver; directed arcs are
linkages that survive Benjamini--Hochberg control ($q<0.05$) in a
Dumitrescu--Hurlin panel non-causality test on a balanced panel of
\NetCountries\ countries (first differences, lag order $K=2$). Sign is
double-encoded: solid blue arcs are synergies, dashed red arcs trade-offs
(cross-country median effect). Arc width is proportional to the standardized
$\tilde{Z}$ statistic, with the strongest 30 arcs drawn with arrowheads at full
opacity and the remainder faded (all \NetEdges\ are listed in
Supplementary Table~\ref*{tab:edges}). Node size is proportional to the absolute value of net
influence; node color runs from purple (net driver) to orange (net receiver),
and the small colored square beside each label is the goal's official UN SDG
identity color. (\textbf{b})~The ten strongest linkages by $\tilde{Z}$, named
and colored by sign. Goal labels follow a single canonical naming scheme used
throughout the figures and tables.}
\label{fig:network}
\end{figure}

\begin{figure}[htbp]
\centering
\includegraphics[width=\textwidth]{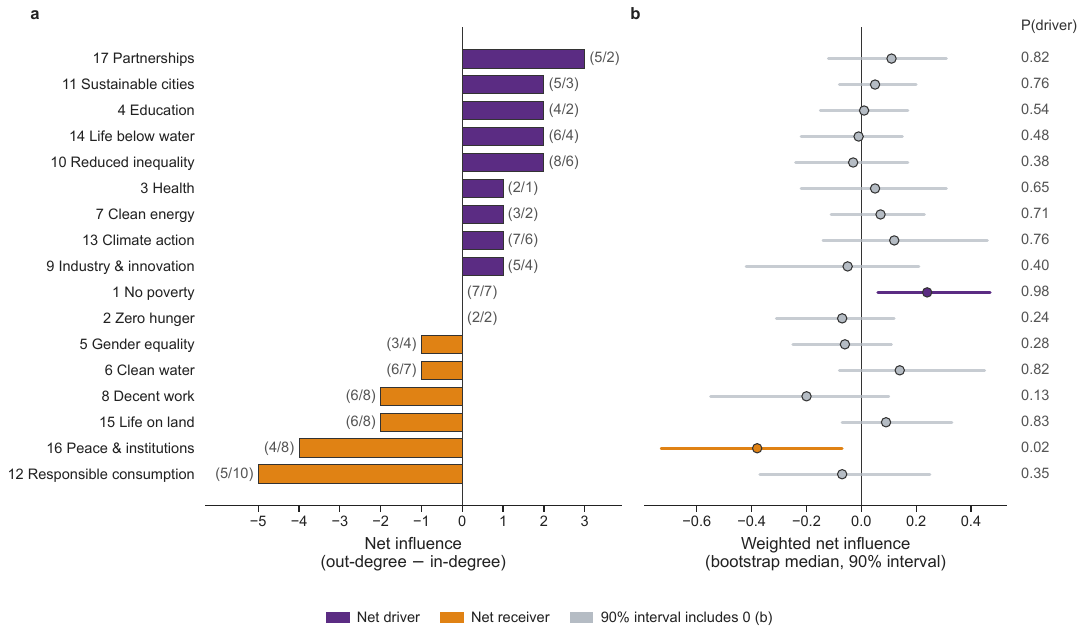}
\caption{Net influence of each goal under two metrics. (\textbf{a})~Raw degree:
out-degree minus in-degree, with (out/in) counts at the bar tips; net drivers
in purple, net receivers in orange. (\textbf{b})~Effect-size-weighted net
influence: country-bootstrap medians with 90\% intervals ($B=\NetBootB$) and,
in the right margin, each goal's bootstrap probability of being a net driver.
Grey marks intervals that include zero; only Peace and Institutions (receiver)
and No Poverty (driver under the weighted metric) are sign-robust. Goals keep
panel~(a)'s order, so disagreement between the panels reads directly as metric
dependence.}
\label{fig:netinf}
\end{figure}

Across the \NetPairs\ directed goal pairs, \NetEdges\ linkages survive false-discovery control
($q<0.05$), giving a directed network of density \NetDensity\
(Figure~\ref{fig:network}). Because the tests share countries, goals, and
common shocks, we also apply the dependence-robust Benjamini--Yekutieli bound:
\NetBYShare\% of the edges (\NetBYEdges\ of \NetEdges) survive it as a strict
subset, the network stays dense (density \NetBYDensity), and the
synergy/trade-off balance is preserved (\NetBYSynergy\ synergies,
\NetBYTradeoff\ trade-offs), so the structural findings do not depend on the
choice of multiplicity correction (edges flagged in Supplementary Table~\ref*{tab:edges}).
The structure is neither sparse nor uniform. Most
goals both send and receive influence, but their balance of the two differs
sharply, and that asymmetry is the policy-relevant signal. Ranking goals by net
influence (out-degree minus in-degree) separates net drivers, whose
improvement Granger-predicts movement in many other goals, from net receivers,
which mostly absorb influence from the rest of the system
(Figure~\ref{fig:netinf}; Table~\ref{tab:centrality}).

\begin{table}[t]
\begin{singlespace}
\centering
\footnotesize
\setlength{\tabcolsep}{4pt}
\renewcommand{\arraystretch}{1.15}
\caption{Goal-level centrality in the directed Granger network under two metrics: raw degree (out-degree, in-degree, net) and standardized-effect-size weighting (country-bootstrap median net influence with 90\% interval, and the bootstrap probability of being a net driver; $B=500$). Bold marks weighted intervals that exclude zero.}
\label{tab:centrality}
\begin{tabular}{lrrrcc}
\toprule
\textbf{Goal} & \textbf{Out} & \textbf{In} & \textbf{Net} & \textbf{Weighted net [90\% CI]} & $\boldsymbol{P}$\textbf{(driver)} \\
\midrule
\multicolumn{6}{@{}l}{\textit{Net drivers (degree metric)}} \\
17 Partnerships & 5 & 2 & $+3$ & $+0.11$~[$-0.12$, $+0.31$] & $0.82$ \\
11 Sustainable cities & 5 & 3 & $+2$ & $+0.05$~[$-0.08$, $+0.20$] & $0.76$ \\
4 Education & 4 & 2 & $+2$ & $+0.01$~[$-0.15$, $+0.17$] & $0.54$ \\
10 Reduced inequality & 8 & 6 & $+2$ & $-0.03$~[$-0.24$, $+0.17$] & $0.38$ \\
14 Life below water & 6 & 4 & $+2$ & $-0.01$~[$-0.22$, $+0.15$] & $0.48$ \\
13 Climate action & 7 & 6 & $+1$ & $+0.12$~[$-0.14$, $+0.46$] & $0.76$ \\
3 Health & 2 & 1 & $+1$ & $+0.05$~[$-0.22$, $+0.31$] & $0.65$ \\
9 Industry \& innovation & 5 & 4 & $+1$ & $-0.05$~[$-0.42$, $+0.21$] & $0.40$ \\
7 Clean energy & 3 & 2 & $+1$ & $+0.07$~[$-0.11$, $+0.23$] & $0.71$ \\
\addlinespace
\multicolumn{6}{@{}l}{\textit{Balanced}} \\
1 No poverty & 7 & 7 & $+0$ & {\boldmath $+0.24$~[$+0.06$, $+0.47$]} & $0.98$ \\
2 Zero hunger & 2 & 2 & $+0$ & $-0.07$~[$-0.31$, $+0.12$] & $0.24$ \\
\addlinespace
\multicolumn{6}{@{}l}{\textit{Net receivers (degree metric)}} \\
5 Gender equality & 3 & 4 & $-1$ & $-0.06$~[$-0.25$, $+0.11$] & $0.28$ \\
6 Clean water & 6 & 7 & $-1$ & $+0.14$~[$-0.08$, $+0.45$] & $0.82$ \\
8 Decent work & 6 & 8 & $-2$ & $-0.20$~[$-0.55$, $+0.10$] & $0.13$ \\
15 Life on land & 6 & 8 & $-2$ & $+0.09$~[$-0.07$, $+0.33$] & $0.83$ \\
16 Peace \& institutions & 4 & 8 & $-4$ & {\boldmath $-0.38$~[$-0.73$, $-0.07$]} & $0.02$ \\
12 Responsible consumption & 5 & 10 & $-5$ & $-0.07$~[$-0.37$, $+0.25$] & $0.35$ \\
\bottomrule
\end{tabular}
\par\smallskip\footnotesize\textit{Notes:} Out-degree is the number of goals a goal Granger-drives; in-degree the number that drive it; net influence is their difference over Benjamini--Hochberg-significant edges ($q<0.05$). The weighted metric sums standardized edge effect sizes ($|$median $\sum_k \beta_k|$ scaled by the source/target first-difference SDs); intervals and $P$(driver) come from resampling countries with replacement. Only Peace/Institutions (receiver) and No Poverty (driver, weighted metric only) have intervals excluding zero: goal-level rankings are metric-dependent.
\end{singlespace}
\end{table}

Under the conventional degree metric, the clearest net drivers are
Partnerships (SDG~17), Sustainable Cities (SDG~11), and Education (SDG~4),
together with Reduced Inequality (SDG~10) and Life Below Water (SDG~14); the
strongest net receivers are Responsible Consumption (SDG~12) and Peace and
Institutions (SDG~16), while No Poverty (SDG~1) is balanced. This ranking
should be read as a description of the estimated network, not as a stable
property of the goal system: as the next subsections show, it survives
standard cross-sectional-dependence controls, but not
changes of lag order or of centrality metric, and bootstrap intervals for most
goals straddle zero (Table~\ref{tab:centrality}, Figure~\ref{fig:netinf}B).

\subsection{Synergies and trade-offs are roughly balanced}

\begin{figure}[htbp]
\centering
\includegraphics[width=\textwidth]{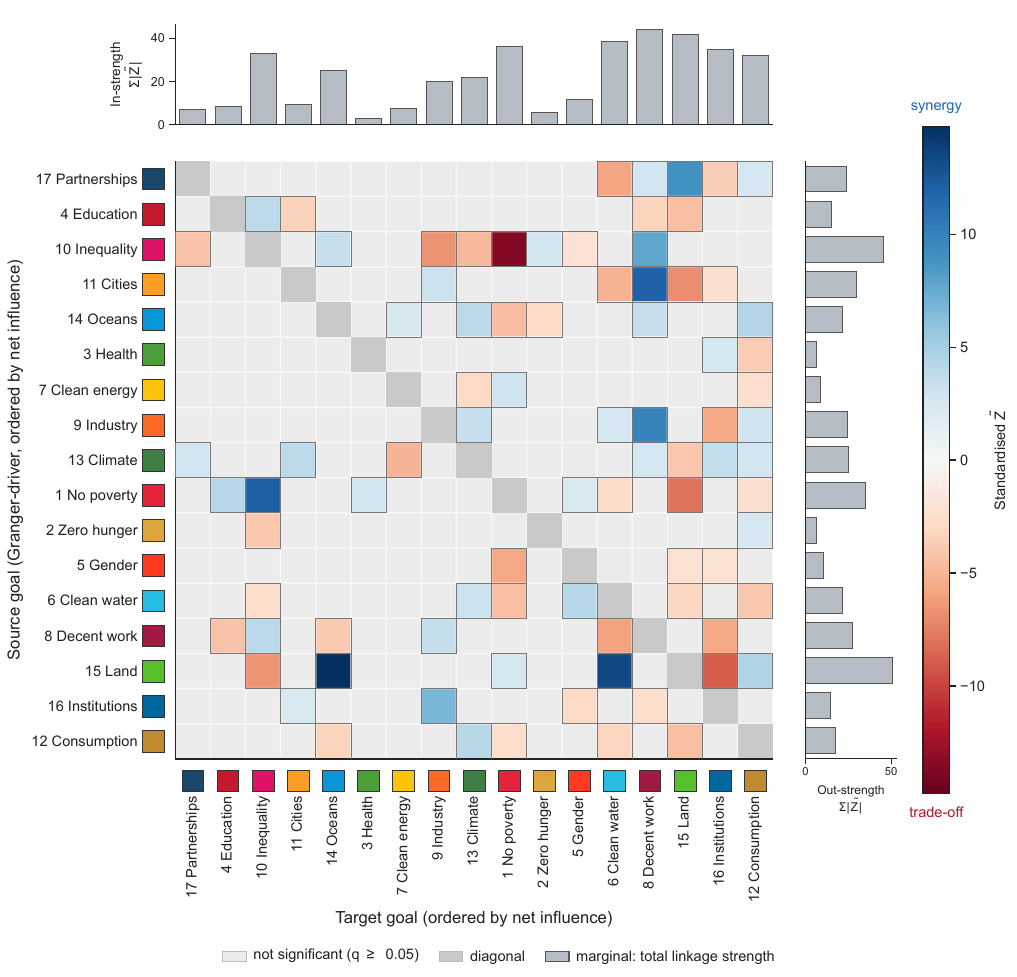}
\caption{Signed adjacency matrix of Benjamini--Hochberg-significant directed
Granger linkages. Rows are source (driver) goals, columns are target goals,
both ordered by net influence; each significant cell is shaded by its
standardized strength and sign (blue: synergy; red: trade-off). Non-significant
pairs are light grey and the diagonal mid grey. Marginal bars give the total
linkage strength ($\Sigma|\tilde{Z}|$) entering each target column (top) and
leaving each source row (right); the colored squares along the axes are the
goals' official UN SDG identity colors.}
\label{fig:heatmap}
\end{figure}

Of the \NetEdges\ retained linkages, \NetSynergy\ are synergies and \NetTradeoff\ are trade-offs, a near
balance that argues against treating SDG interdependence as uniformly
reinforcing. The strongest synergies sit among the environmental goals, where
Life on Land and Life Below Water co-improve and Life on Land Granger-predicts
Clean Water, and within the prosperity cluster, where No Poverty and Reduced
Inequality reinforce one another and Sustainable Cities and Industry both
predict Decent Work and Growth. The strongest trade-offs involve tension between
distributional and aggregate goals: improvements in measured Reduced Inequality
are followed by slower movement in the No Poverty score, and land-use gains
trade against several socioeconomic goals. Because synergies and trade-offs
coexist at comparable strength, a single goal cannot be designated a universal
accelerator. The useful question is which \emph{directed} linkages a given
country can exploit, not which goal to maximize in isolation. The complete
signed structure, which source goal drives which target and with what sign, is
shown as a source-by-target matrix in Figure~\ref{fig:heatmap}.

\subsection{How robust is the driver--receiver ranking?}

Interlinkage studies typically report a driver ranking and probe its
stability against at most a small set of alternative specifications
\citep{pradhan2017systematic,lusseau2019income,asadikia2024navigating}. We
instead stress-test
the ranking against every dimension a skeptical reader could vary, and the
answer is asymmetric: the existence and density of directed structure is
robust, the goal-level hierarchy is not.

\begin{figure}[htbp]
\centering
\includegraphics[width=\textwidth]{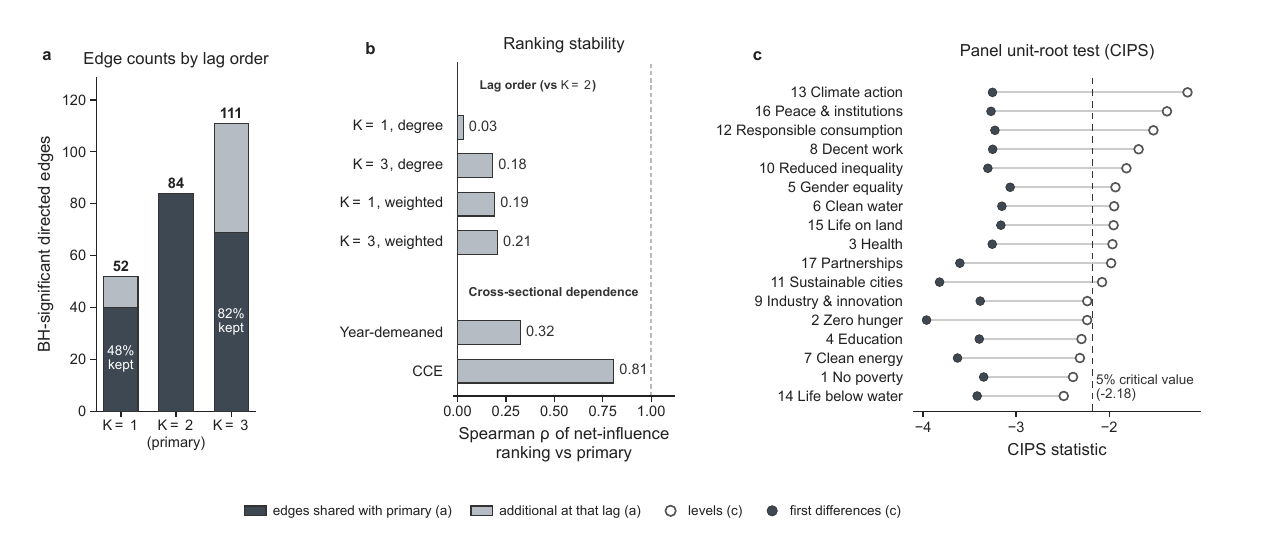}
\caption{Robustness of the directed network. (\textbf{a})~Lag-order
sensitivity: at each lag order $K=1,2,3$ the bar splits into edges shared with
the primary ($K=2$) network (dark) and edges additional at that lag (light),
with the retained share annotated; sign agreement on shared edges is
\NetLagThreeSign\% at $K=3$ (72\% at $K=1$). (\textbf{b})~Ranking stability:
Spearman correlation of the goal net-influence ranking with the primary network,
grouped into lag-order variations (degree and effect-size-weighted metrics) and
cross-sectional-dependence controls; edge membership is moderately stable while
rankings are not, except conditionally under CCE. (\textbf{c})~Pesaran (2007)
CIPS panel unit-root statistics for each goal composite in levels (open circles)
and first differences (filled); the dashed line is the 5\% critical value
(\CIPScv). Levels cluster near or above the threshold (unit-root behavior),
whereas all composites are stationary in first differences.}
\label{fig:robust}
\end{figure}

Three results establish the fragility. First, lag order: the Spearman
correlation between the primary ($K=2$) net-influence ranking and its $K=1$
and $K=3$ counterparts is \NetDegSpearKOne\ and \NetDegSpearKThree\ under the
degree metric, and \NetWtSpearKOne\ and \NetWtSpearKThree\ under
effect-size weighting (Figure~\ref{fig:robust}B), even though edge
\emph{membership} overlaps substantially at $K=3$
(\NetLagThreeOverlap\%; Figure~\ref{fig:robust}A). Second, metric: degree and
effect-size-weighted centrality disagree about individual goals. Education is
a net driver by degree but its bootstrap probability of being a weighted net
driver is only \PDrvEducation; Responsible Consumption, the strongest degree
receiver, is ambiguous under weighting ($P(\text{driver})=\PDrvConsumption$,
where a robust receiver would sit near zero). Third,
uncertainty: across \NetBootB\ country-bootstrap resamples, 90\% intervals for
weighted net influence include zero for fifteen of seventeen goals
(Table~\ref{tab:centrality}). The exceptions are informative. Peace and
Institutions is a net receiver under every metric, lag order, and resample
($P(\text{driver})=\PDrvPeace$; interval [\PeaceCIlow, \PeaceCIhigh]), and
No Poverty, balanced by degree, emerges as the most probable weighted net
driver ($P=\PDrvNoPoverty$; [\NoPovertyCIlow, \NoPovertyCIhigh]). Partnerships
($P=\PDrvPartnerships$) and Sustainable Cities ($P=\PDrvCities$) lean toward
driver status under both metrics without reaching significance.

The cross-sectional-dependence controls now standard in macro-panel work
\citep{pesaran2006cce,pesaran2015cd} do not rescue the ranking so much as
scope it. Conditional on the degree metric
and $K=2$, the ranking is robust to CSD: under the common correlated effects
estimator it correlates at \NetCCESpearman\ with the primary and
\NetCCECore\ of \NetEdges\ primary edges survive (Figure~\ref{fig:robust}B).
Sub-period estimates point the same way as the lag and metric checks: dense
connectivity and the synergy/trade-off balance appear in every estimable
window, while edge identity and rankings shift across the pre-SDG, SDG-era,
and pre-COVID windows (retained shares \SubRetainedPreSDG--%
\SubRetainedPreCovid\%; Supplementary Table~\ref*{tab:subperiod}). Finally, edge signs
themselves average over substantial country heterogeneity: the median
cross-country sign consensus is \HetMedianConsensus\%
(Supplementary Table~\ref*{tab:edges}), exactly what the Dumitrescu--Hurlin alternative
(causality for \emph{some} countries) anticipates. The constructive
conclusion is that goal-level league tables are the wrong policy object in
this system; the durable objects are the dense directed structure, its
synergy/trade-off balance, and the specific supported pathways below.

\begin{table}[htbp]
\begin{singlespace}
\centering
\scriptsize
\setlength{\tabcolsep}{3pt}
\renewcommand{\arraystretch}{1.12}
\caption{Strongest directed Granger linkages in the SDG network. Rows reproduce the top entries from the full edge catalog in Supplementary Table~\ref*{tab:edges}. BY marks linkages that also survive the dependence-robust Benjamini--Yekutieli bound; consensus is the share of countries whose summed lag coefficient has the median sign.}
\label{tab:edge_excerpt}
\begin{tabularx}{\textwidth}{@{}>{\raggedright\arraybackslash}p{0.20\textwidth}>{\raggedright\arraybackslash}p{0.20\textwidth}rccr>{\raggedleft\arraybackslash}p{0.08\textwidth}@{}}
\toprule
\textbf{Source goal} & \textbf{Target goal} & $\boldsymbol{\tilde{Z}}$ & $\boldsymbol{q}$ & \textbf{BY} & \textbf{Sign} & \textbf{Cons. \%} \\
\midrule
15 Life on land & 14 Life below water & 14.77 & $<$0.001 & \checkmark & Syn. & 58 \\
10 Reduced inequality & 1 No poverty & 13.58 & $<$0.001 & \checkmark & Trade. & 68 \\
15 Life on land & 6 Clean water & 13.27 & $<$0.001 & \checkmark & Syn. & 54 \\
1 No poverty & 10 Reduced inequality & 12.09 & $<$0.001 & \checkmark & Syn. & 64 \\
11 Sustainable cities & 8 Decent work & 11.91 & $<$0.001 & \checkmark & Syn. & 50 \\
9 Industry \& innovation & 8 Decent work & 9.85 & $<$0.001 & \checkmark & Syn. & 55 \\
15 Life on land & 16 Peace \& institutions & 8.96 & $<$0.001 & \checkmark & Trade. & 53 \\
17 Partnerships & 15 Life on land & 8.93 & $<$0.001 & \checkmark & Syn. & 56 \\
1 No poverty & 15 Life on land & 7.99 & $<$0.001 & \checkmark & Trade. & 51 \\
10 Reduced inequality & 8 Decent work & 7.66 & $<$0.001 & \checkmark & Syn. & 50 \\
11 Sustainable cities & 15 Life on land & 6.86 & $<$0.001 & \checkmark & Trade. & 54 \\
16 Peace \& institutions & 9 Industry \& innovation & 6.72 & $<$0.001 & \checkmark & Syn. & 57 \\
10 Reduced inequality & 9 Industry \& innovation & 6.62 & $<$0.001 & \checkmark & Trade. & 54 \\
15 Life on land & 10 Reduced inequality & 6.52 & $<$0.001 & \checkmark & Trade. & 57 \\
8 Decent work & 6 Clean water & 5.96 & $<$0.001 & \checkmark & Trade. & 60 \\
\bottomrule
\end{tabularx}
\parbox{\textwidth}{\footnotesize \textit{Notes:} Sign labels summarize the cross-country median summed lag coefficient, not a universal country-level response. Moderate consensus is expected under the Dumitrescu--Hurlin alternative, which detects Granger predictability for some countries.}
\end{singlespace}
\end{table}

\subsection{Dynamic linkages build over several years}

\begin{figure}[htbp]
\centering
\includegraphics[width=\textwidth]{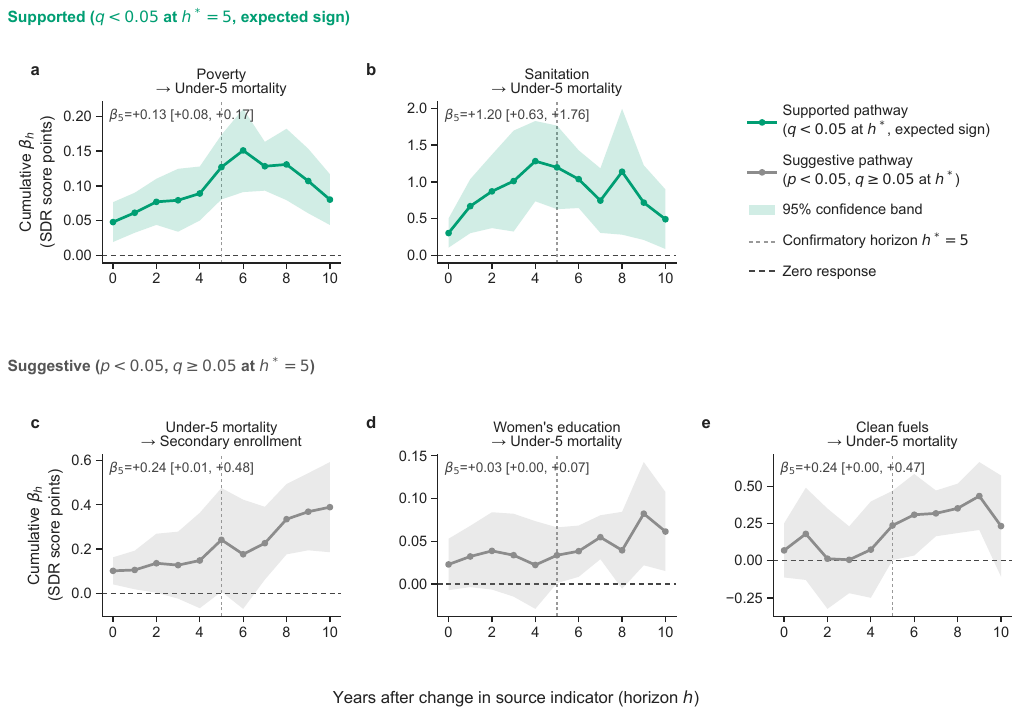}
\caption{Dynamic responses of theory-derived SDG linkages, estimated by panel
local projections with country and year fixed effects and Driscoll--Kraay
standard errors. Each panel plots the cumulative response $\beta_h$ (SDR score
points) to a one-point improvement in the source indicator over horizons
$h=0$--$10$ years, with 95\% confidence bands; the horizontal dashed line marks
the null and the vertical dashed line the prespecified confirmatory horizon
$h^*=5$. The top row holds the two supported linkages (Benjamini--Hochberg
$q<0.05$ at $h^*$ in the expected direction, green); the bottom row the three
suggestive linkages ($p<0.05$, $q\geq0.05$, grey). Panels (\textbf{a})--%
(\textbf{e}) read left to right, top to bottom; the spare cell in the top row
is the key.}
\label{fig:irf}
\end{figure}

The panel local projections show that the prespecified indicator linkages, where
present, act with multi-year lags rather than contemporaneously
(Figure~\ref{fig:irf}). Judged by the cumulative response at the prespecified
\LPHorizonStar-year horizon under false-discovery control, \LPSupported\ of the
estimable linkages are supported. Both run into child survival: a one-point
improvement in sanitation access raises the under-five-mortality score by about
1.2 points over five years ($q<0.001$), and poverty reduction predicts lower
under-five mortality over the same horizon ($q<0.001$). Each builds steadily
rather than appearing immediately, peaking around four to five years
(Supplementary Table~\ref*{tab:lp}).

A further \LPSuggestive\ linkages are suggestive (raw $p<0.05$ at the
\LPHorizonStar-year horizon but not surviving false-discovery control):
reductions in under-five mortality predicting later gains in secondary
education, and women's education and clean cooking fuel each predicting lower
child mortality. The remaining linkages are unsupported or, where the
constituent series have too little variation, not interpretable; the complete
31-pair audit is reported in Table~\ref{tab:lpfull}, with a null exemplar
shown in Supplementary Figure~S1. The robust qualitative
finding is that the surviving SDG linkages are dynamic, accruing over a four-
to five-year window, so cross-sectional or single-lag analyses will tend to
understate them.

Timing and feedback diagnostics qualify how the two supported pathways should
be read (Supplementary Table~\ref*{tab:placebo}). Annual changes in these slow-moving
indicators are persistent, so lead placebos are significant for both pathways,
as expected; the informative tests are the joint horse race and the reverse
direction. The poverty pathway retains its current-change coefficient when the
lead is included, but its reverse-direction projection is also significant
($\beta_5=\PovRevBeta$, $p=\PovRevP$): poverty reduction and child survival
form a reinforcing loop rather than a one-way lever. The sanitation pathway
shows no reverse feedback ($p=\SanRevP$), but its joint horse race favors
the lead term over the current change, a pattern consistent with smoothing or
timing misalignment in the underlying measurement; we therefore read it as a
robust predictive association whose precise timing cannot be resolved in
annual differences. Both pathways hold with the same sign and significance in both
income strata, with larger sanitation effects in poorer countries
(Supplementary Table~\ref*{tab:strata}, Supplementary Figure~S2).

\begin{table}[p]
\centering
\begingroup
\fontsize{5.8}{6.2}\selectfont
\setlength{\tabcolsep}{1.6pt}
\renewcommand{\arraystretch}{0.92}
\caption{Complete local-projection results for all 31 prespecified indicator linkages at the confirmatory five-year horizon. Exp.\ is the hypothesized sign; pairs are not interpretable when the balanced pair panel falls below 30 countries or the confirmatory regression is not estimable.}
\label{tab:lpfull}
\begin{tabularx}{\textwidth}{@{}>{\raggedright\arraybackslash}Xccr>{\raggedright\arraybackslash}p{0.20\textwidth}cc>{\raggedright\arraybackslash}p{0.14\textwidth}@{}}
\toprule
Linkage & Cl. & Exp. & $N$ & $\beta_5$ [95\% CI] & $p$ & $q$ & Status \\
\midrule
Poverty $\rightarrow$ Secondary enrollment & A & $+$ & $154$ & $+0.018$~[$-0.06$, $+0.10$] & $0.638$ & $0.744$ & unsupported \\
Poverty $\rightarrow$ Under-5 mortality & A & $+$ & $159$ & $+0.127$~[$+0.08$, $+0.17$] & $<$0.001 & $<$0.001 & supported \\
Secondary enrollment $\rightarrow$ GDP growth & A & $+$ & $178$ & -- & -- & -- & not interpretable \\
Under-5 mortality $\rightarrow$ Secondary enrollment & A & $+$ & $184$ & $+0.242$~[$+0.01$, $+0.48$] & $0.043$ & $0.137$ & suggestive \\
Under-5 mortality $\rightarrow$ GDP growth & A & $+$ & $185$ & -- & -- & -- & not interpretable \\
GDP growth $\rightarrow$ Poverty & A & $+$ & $155$ & -- & -- & -- & not interpretable \\
Women's education $\rightarrow$ Under-5 mortality & B & $+$ & $190$ & $+0.034$~[$+0.00$, $+0.07$] & $0.043$ & $0.137$ & suggestive \\
Women's education $\rightarrow$ Undernourishment & B & $+$ & $179$ & $-0.017$~[$-0.19$, $+0.16$] & $0.850$ & $0.902$ & unsupported \\
Secondary enrollment $\rightarrow$ Female labor force & B & $+$ & $170$ & $-0.010$~[$-0.03$, $+0.01$] & $0.260$ & $0.419$ & unsupported \\
Female labor force $\rightarrow$ GDP growth & B & $+$ & $172$ & -- & -- & -- & not interpretable \\
Drinking water $\rightarrow$ Under-5 mortality & C & $+$ & $192$ & $+0.653$~[$-0.78$, $+2.09$] & $0.372$ & $0.517$ & unsupported \\
Sanitation $\rightarrow$ Under-5 mortality & C & $+$ & $192$ & $+1.197$~[$+0.63$, $+1.76$] & $<$0.001 & $<$0.001 & supported \\
Drinking water $\rightarrow$ Undernourishment & C & $+$ & $181$ & $-3.415$~[$-6.18$, $-0.65$] & $0.016$ & $0.082$ & unsupported \\
Sanitation $\rightarrow$ Stunting & C & $+$ & $187$ & $+1.180$~[$-1.17$, $+3.53$] & $0.324$ & $0.486$ & unsupported \\
Corruption $\rightarrow$ GDP growth & D & $+$ & $172$ & -- & -- & -- & not interpretable \\
Corruption $\rightarrow$ Secondary enrollment & D & $+$ & $170$ & $+0.149$~[$-0.01$, $+0.31$] & $0.062$ & $0.146$ & unsupported \\
Homicides $\rightarrow$ GDP growth & D & $+$ & $161$ & -- & -- & -- & not interpretable \\
GDP growth $\rightarrow$ CO$_2$ emissions & E & $-$ & $184$ & -- & -- & -- & not interpretable \\
CO$_2$ emissions $\rightarrow$ Air pollution & E & $+$ & $190$ & $+0.000$~[$-0.07$, $+0.07$] & $0.999$ & $0.999$ & unsupported \\
CO$_2$ emissions $\rightarrow$ Slums & E & $+$ & $183$ & $+0.095$~[$-0.00$, $+0.19$] & $0.056$ & $0.146$ & unsupported \\
Electricity $\rightarrow$ GDP growth & E & $+$ & $185$ & -- & -- & -- & not interpretable \\
Electricity $\rightarrow$ Secondary enrollment & E & $+$ & $185$ & $+0.071$~[$-0.02$, $+0.16$] & $0.107$ & $0.224$ & unsupported \\
Clean fuels $\rightarrow$ Under-5 mortality & E & $+$ & $189$ & $+0.237$~[$+0.00$, $+0.47$] & $0.045$ & $0.137$ & suggestive \\
Undernourishment $\rightarrow$ Under-5 mortality & F & $+$ & $180$ & $+0.029$~[$-0.07$, $+0.13$] & $0.562$ & $0.694$ & unsupported \\
Stunting $\rightarrow$ Secondary enrollment & F & $+$ & $179$ & $+0.027$~[$-0.02$, $+0.07$] & $0.254$ & $0.419$ & unsupported \\
GDP growth $\rightarrow$ Undernourishment & F & $+$ & $174$ & -- & -- & -- & not interpretable \\
UHC $\rightarrow$ Under-5 mortality & G & $+$ & $192$ & $-0.091$~[$-0.15$, $-0.03$] & $0.002$ & $0.011$ & unsupported \\
UHC $\rightarrow$ Life expectancy & G & $+$ & $192$ & $-0.025$~[$-0.08$, $+0.03$] & $0.394$ & $0.517$ & unsupported \\
Income inequality $\rightarrow$ Under-5 mortality & H & $+$ & $163$ & $-0.017$~[$-0.04$, $+0.01$] & $0.231$ & $0.419$ & unsupported \\
Income inequality $\rightarrow$ Secondary enrollment & H & $+$ & $159$ & $+0.005$~[$-0.06$, $+0.07$] & $0.859$ & $0.902$ & unsupported \\
GDP growth $\rightarrow$ Income inequality & H & $+$ & $160$ & -- & -- & -- & not interpretable \\
\bottomrule
\end{tabularx}
\endgroup
\end{table}
\FloatBarrier

\subsection{The education--health pathway holds in independent data}

The reinforcing education--health linkage that appears in both the network and
the local projections is not an artifact of the SDR construction. In
independent World Development Indicators data covering \WdiCountries\
countries, a lagged increase in gross secondary enrollment predicts a
subsequent reduction in under-five mortality ($\hat{\theta}=\WdiBeta$
[\WdiCIlow, \WdiCIhigh] deaths per 1{,}000 live births per percentage point,
$p<0.001$, Supplementary Table~\ref*{tab:wdi}; about \WdiPerTenPP\ per 10-percentage-point
enrollment rise). The association is robust to controls for government health
expenditure from the WHO Global Health Expenditure Database
\citep{who2026gheddataset}, education expenditure, GDP per capita, and fertility,
and stable in the pre-2020
sub-sample. It attenuates under country-specific trends and within the
near-ceiling balanced-panel subsample (Supplementary Table~\ref*{tab:wdi}), so we read it as
an out-of-source agreement on direction: a monitorable association, not an
identified intervention effect.

\section{Policy Implications: From Rankings to Directed Portfolios}


\subsection{From rankings to linkage-level evidence}

The directed network reframes the prioritization problem, but not by handing
governments a better league table. Our robustness analysis shows that
goal-level driver rankings, ours included, shift with the lag order and the
centrality metric, and that bootstrap intervals straddle zero for most goals.
The defensible reading is narrower and more useful. Peace and Institutions is
robustly a net receiver, sustained by progress elsewhere rather than an
efficient place to push for system-wide gains; Partnerships and Sustainable
Cities lean consistently toward driver status; and under the
effect-size-weighted metric poverty reduction is the most probable net driver,
which cautions against treating it purely as a downstream outcome. For a
government with limited capacity, the operational unit of evidence should be
the specific directed linkage, with its sign, horizon, and supporting
diagnostics, not a goal's position in a ranking.

We read this reframing as itself a policy-relevant finding. Accelerator lists
derived from unweighted interlinkage counts populate much of the advisory
literature \citep{un2023gsdr,allen2019prioritising,asadikia2024navigating},
and the same fragility we document here applies to them. Net influence is an
average over a heterogeneous panel, the network contains both synergies and
trade-offs in comparable numbers, edge signs summarize substantial
cross-country heterogeneity, and individual countries face their own
structure. The global map is a prior for where transmission is most likely,
to be combined with local evidence. It is not a fixed allocation rule.

\subsection{Manage trade-offs, not only synergies}

Of the \NetEdges\ retained linkages, \NetTradeoff\ are trade-offs and \NetSynergy\ are synergies. A portfolio
that pursues synergies while ignoring trade-offs will encounter unanticipated
tensions. The data show
recurring trade-offs between distributional and aggregate goals (improvements
in measured inequality are followed by slower movement in the poverty score)
and between land use and several socioeconomic goals. Coordinated planning
should therefore pair driver investments with explicit monitoring of the goals
most likely to move against them, so that a synergy in one direction is not
bought at an unmonitored cost in another. Reporting interdependence as a
balanced map of synergies and trade-offs represents the evidence more
faithfully than a synergy-only narrative, and gives planners the information
a portfolio decision actually requires.

\subsection{Plan for multi-year transmission}

The supported dynamic linkages act with lags of four to five years rather than
within a single budget cycle. Sanitation improvements and poverty reduction each
raise child-survival scores over roughly four to five years. Two implications
follow. First, review dates for cross-goal mechanisms should be set years, not
months, after an intervention; judging a linkage at a one-year horizon will
usually understate it. Second, monitoring should track the \emph{trajectory} of
the expected response over several years, not a single follow-up reading.

\subsection{Use the supported linkages as monitoring hypotheses}

Two pathways are robust enough across the reported diagnostics to anchor
specific monitoring hypotheses. Both concern child survival. Where a portfolio expands sanitation access, or reduces
poverty, planners can treat improvement in under-five mortality over the
subsequent four to five years as the expected and monitorable response, in
both lower- and higher-income settings. The poverty pathway runs in both
directions, since child survival also predicts later poverty reduction, so it
is best planned as a reinforcing loop in which gains compound, rather than a
one-way lever. A reinforcing relationship between education and child health
is a weaker (suggestive) hypothesis: schooling and child survival appear to
move together over half a decade, and the education-to-child-mortality
direction is corroborated in independent World Development Indicators data.
These relationships are plausible through several channels, among them
sanitation-borne disease reduction \citep{wolf2023burden}, household
resources, women's agency, fertility timing, and service use
\citep{gakidou2010education}. The present design does not separate these
channels, and the estimates are Granger-predictive rather than intervention
effects. Programs should therefore treat them as priors to monitor, measure
the candidate mechanisms rather than assume them, and not treat any single
upstream goal as a substitute for direct provision in the downstream sector.

\subsection{A decision protocol for governments and development partners}

We distill the evidence into a five-step protocol. First, define the decision at the
indicator and population level: ``accelerate SDG~4'' is too broad, whereas
``increase secondary-school retention among rural girls while maintaining
learning quality'' identifies an outcome and a population. Second, choose entry
points from supported directed linkages rather than from a goal ranking, and
state the linkage, with its expected sign and multi-year horizon, that connects
the action to downstream outcomes.
Third, build a portfolio containing both the upstream action and direct measures
for the downstream outcome, and identify the trade-off goals to watch. Fourth,
specify leading and lagging indicators, expected direction, and a review date
several years out. Fifth, adapt the portfolio if the mechanism indicators do not
move along the expected trajectory.

This protocol uses global evidence as a prior, not a universal rule. The directed
network indicates which mechanisms are most likely to transmit; national
administrative records, household surveys, and implementation studies update
those priors. Cross-ministry coordination is then organized around jointly
monitored outcomes and horizons rather than ownership of an abstract accelerator
goal: education and health agencies can agree on schooling, child-health, and
service-use indicators, and development partners can align support with the same
directed map.

\subsection{Implications for SDG reporting}

SDG dashboards \citep{sachs2025sdr} communicate composite progress well but
obscure directionality and timing. When policy is justified by an interlinkage, reporting should display the
direction of the claimed linkage, its expected horizon, and the constituent
indicators on both ends. Rather than a single accelerator ranking, agencies can
publish a directed map with three tiers: linkages supported in the relevant or
comparable sample; linkages plausible but only suggestive; and relationships that
fail diagnostic or stability requirements. This is more informative than an
undirected, uncorrected graph, shows where investment in data is needed, and
reduces the risk that an exploratory edge becomes a permanent policy slogan.
\section{Discussion}


\subsection{What a directed, dynamic map adds}

Mapping the goal system as a directed, dynamically resolved network changes what
the interlinkage literature can offer policy. Undirected correlation and
expert-coded maps establish that goals are connected; they do not say which goals
drive others or how long transmission takes. The analysis here recovers
direction under false-discovery control, quantifies the response horizon of
each supported linkage, and then stress-tests the goal-level summary
statistics built on top of the network. This converts a connectivity claim
into something more disciplined: a dense directed structure whose reliable
policy content lives at the level of specific linkages, not goal rankings.

This framing also reconciles apparently conflicting SDG studies. Dense networks
and sparse directed networks arise from the same data because they answer
different questions and use different filters. An uncorrected, level-based
network counts every visible association, including those generated by shared
trends among persistent series. A directed network estimated on
stationary-transformed data and screened for false discovery counts only linkages
that survive both filters. The retained network here, \NetEdges\ of \NetPairs\ possible
directed edges, is the second kind: sparser than a raw graph, but each edge is
eligible for interpretation.

\subsection{Drivers, receivers, and the balance of synergies and trade-offs}

The baseline degree ranking has substantive appeal, with enabling goals such
as partnerships, cities, and education at the top and outcome goals such as
institutions and responsible consumption at the bottom, and it echoes a long
tradition in development thinking that treats schooling, infrastructure, and
urban systems as upstream of broad welfare gains
\citep{barro1991economic,sen1999development}. Our robustness analysis
counsels against leaning on that appeal. The ranking's correlation with
itself across lag orders is close to zero, degree-based and
effect-size-weighted centrality disagree about individual goals (education
most prominently), and country-bootstrap intervals straddle zero for fifteen
of seventeen goals. What survives is more specific. Peace and Institutions is
the system's robust net receiver; institutional quality, in these data, is
sustained by progress elsewhere rather than predicting it. Poverty
reduction, balanced by degree, is the most probable weighted driver, embedded
in a reinforcing loop with child survival. We regard the fragility result as
a contribution in its own right: driver hierarchies built on unweighted edge
counts, which remain common in the interlinkage literature and its policy
translations \citep{pradhan2017systematic,asadikia2024navigating,un2023gsdr},
should be treated as descriptive summaries of one
specification, not as evidence about which goals to fund first.

The near balance of synergies (\NetSynergy) and trade-offs (\NetTradeoff) is equally important. SDG
interdependence is frequently presented as predominantly reinforcing
\citep{pradhan2017systematic,phamtruffert2020interactions}; in these
data, improvements in one goal are about as likely to be followed by tension in
another as by co-improvement. The strongest trade-offs sit between distributional
and aggregate goals and between land use and socioeconomic outcomes. On these
estimates, treating interdependence as uniformly synergistic would overstate
how easily gains compound, and would hide the tensions a portfolio must
manage.

\subsection{Why estimators matched to the panel matter}

The results depend on using estimators suited to a wide, moderately long,
trending, cross-sectionally dependent panel. Short-panel GMM
\citep{arellano1991some,blundell1998initial}, often applied to
such data, over-instruments at $T=25$ \citep{roodman2009note} and produces
unstable or degenerate
estimates; it is the wrong tool for this regime rather than a benchmark the data
fail. Conducting inference on series levels is equally hazardous: persistent,
co-trending SDG scores generate spurious causality almost everywhere
\citep{granger1974spurious}, so a
level-based Granger network is dense and largely uninformative. The CIPS test
confirms the diagnosis. The unit-root null is not rejected in levels for
\CipsLevelUnitRoot\ of \CipsGoals\ composites, yet all are stationary in first
differences (Figure~\ref{fig:robust}C). First-differencing collapses that spurious density to an
interpretable structure, and the Dumitrescu--Hurlin test accommodates the heterogeneous
country dynamics that a pooled model would misrepresent. Local projections with
Driscoll--Kraay standard errors then recover dynamic magnitudes without imposing
the instrument structure that disables GMM. The methodological lesson generalizes
beyond the SDGs: directed-network claims from macro panels should be built on
stationarity-aware, dependence-robust estimators and explicit multiplicity
control.

\subsection{Measurement and triangulation}

Composite goal scores serve legitimate purposes but embed weights, normalization,
and constituent overlap that can shape estimated relationships. The
education-to-child-health linkage appears in the SDR composites, in independent
World Development Indicators data, and in both the network and the local
projections. This convergence is reassuring: the linkage is unlikely to be an
artifact of one construction. Where a composite linkage cannot be reproduced at the indicator
level, by contrast, its policy meaning remains uncertain, which is why the
supported claims are anchored to specific indicators and horizons rather than to
goal scores alone.

\subsection{Relation to current research}

These results are consistent with the literature's movement away from
context-free rankings. Expert frameworks emphasize context \citep{nilsson2018mapping};
empirical networks vary with income and development
\citep{lusseau2019income,wu2022decoupling}; multi-criteria approaches combine
interactions with gaps and institutional priorities
\citep{allen2019prioritising,asadikia2024navigating}; and simulation work favors
portfolios under uncertainty \citep{yang2025portfolios}. Recent reviews call for
more directional, indicator-level, and policy-mechanism evidence
\citep{issa2024network,khot2026gaps}. The present study contributes a directed,
dynamically resolved, false-discovery-controlled map across all seventeen goals,
estimated with methods matched to the panel, with every reported linkage keyed to
a reproducible result registry recording its family, direction, statistic,
multiplicity-adjusted value, sample, and claim status.

\subsection{Limitations}

First, Granger causality \citep{granger1969} is predictive, not structural.
The directed edges and
impulse responses describe which changes precede which, conditional on history
and common time effects; they do not identify intervention effects and can be
confounded by omitted common drivers. The policy reading is accordingly framed as
priors and monitoring hypotheses, not as estimated returns to specific programs.

Second, the balanced \NetCountries-country panel selects countries with
complete data on all seventeen goals for all twenty-five years. The selection
operates through data completeness rather than income, region, or governance
composition (Supplementary Table~\ref*{tab:selection}), and the panel covers \SelPopShare\%
of world population. Even so, the most data-sparse states, which are often
fragile states, are necessarily absent, and conclusions for those settings
are extrapolation.

Third, the annual frequency and 25-year span bound the dynamics that can be
observed. The local projections trace responses to ten years, but very short
implementation cycles and generational horizons lie outside the window, and
structural breaks remain possible despite first-differencing.

Fourth, working in first differences answers a short-to-medium-run question about
year-over-year changes; long-run equilibrium (cointegrating) relationships among
the levels are a complementary question this design does not address. The
cross-country median used to sign each edge summarizes heterogeneous coefficients
and can mask countries that move against the average.

Fifth, lag order and metric choice matter more than is usually acknowledged.
Edge membership is moderately stable between $K=2$ and $K=3$
(\NetLagThreeOverlap\% retained, \NetLagThreeSign\% sign agreement), but only
about half of the edges survive at $K=1$, the lag length per-country
information criteria actually prefer, and goal-level rankings are unstable
across both lags and centrality metrics. Individual edge claims and any
ranking-based reading should be tempered accordingly; the same applies to the
sub-period sensitivity of edge identity and to the moderate cross-country
sign consensus behind each edge label. Sixth,
the \LPPairs\ local-projection linkages are a theory-derived subset, not the full
indicator space; supported status within this set is not a claim about untested
pairs. Finally, global outcome panels omit the institutional,
distributional, and political-economy information that policy portfolios require;
a supported predictive linkage does not establish that a given intervention is
feasible, equitable, or cost-effective.

\subsection{Reproducibility and future research}

Every headline number, table row, and figure is generated from the standardized
result registry, with source manifests recording file hashes and provenance, and
tests enforcing the balanced panel, the false-discovery families, and
deterministic claim classification. Future work should pursue designs that
strengthen interpretation rather than add network complexity: panel cointegration
and error-correction models to recover long-run structure alongside the
short-run dynamics reported here; quasi-experimental or program evaluations of
the supported pathways from sanitation and from poverty to child mortality;
extension of the directed map to the full indicator space and to income- and
region-specific sub-networks; and country-level portfolio studies that combine
these global priors with administrative data, distributional objectives, and
implementation constraints.
\section{Conclusion}

SDG interdependence is dense and directed, but what it offers policy is
carried by specific linkages rather than by goal rankings. In a balanced
panel of \NetCountries\ countries from 2000 to 2024, estimators matched to a
wide, trending, cross-sectionally dependent panel recover a directed network
of \NetEdges\ false-discovery-controlled linkages in which synergies and
trade-offs occur at comparable strength, so no single goal is a universal
accelerator. The goal-level driver--receiver hierarchy is itself fragile.
Rankings correlate weakly across lag orders and centrality metrics, and under
a country bootstrap only two roles are distinguishable from zero: peace and
strong institutions as the system's clearest net receiver, and poverty
reduction as the most probable effect-size-weighted driver, with partnerships
and sustainable cities the most consistent driver candidates short of
significance. What survives every
screen is a small set of dynamic pathways that accrue over four to five
years and hold in both poorer and richer countries: improvements in
sanitation access and in poverty each precede gains in child survival; the
poverty pathway runs in both directions, so its gains compound in a
reinforcing loop; and the association between education
and child health is corroborated in independent data. The defensible policy
response is not a fixed global accelerator rule but a portfolio that anchors
investment cases on supported directed linkages rather than goal-level league
tables, anticipates the trade-offs those linkages entail, and monitors the
constituent indicators along their expected multi-year trajectories.

\FloatBarrier

\bmhead{Funding}
This research did not receive any specific grant from funding agencies in the
public, commercial, or not-for-profit sectors.

\bmhead{Competing interests}
The authors declare no competing interests.

\bmhead{Open Research}
The study uses four public input resources: the Sustainable Development Report
2025 SDG Index and Dashboards data, including the backdated SDG Index workbook,
available through the SDG Transformation Center 2025 data archive and Data Hub
\citep{sdsn2025sdrdata}; World Development Indicators, available through the
World Bank DataBank \citep{worldbank2026wdidataset}; the World Health
Organization Global Health Expenditure Database \citep{who2026gheddataset}; and
World Bank country classification metadata, available through the World Bank
Country API \citep{worldbank2026countrymetadata}. The analysis code is publicly
available in a GitHub repository \citep{fahim2026analysiscode}.

\bmhead{Author contributions}
M.M.M.\ Fahim conceived the study and designed the methodology.
M.M.M.\ Fahim curated the data and performed the formal analysis. M.M.M.\ Fahim,
M.J.H.\ Imran, and L.\ Debnath conducted the investigation. Software was
developed by M.M.M.\ Fahim, T.\ Shil, E.B.\ Pranto, and M.N.\ Molla. Validation
was performed by L.\ Debnath, T.\ Shil, M.M.M.\ Fahim, and M.N.\ Molla.
Visualization was produced by M.M.M.\ Fahim, E.B.\ Pranto, and M.J.H.\ Imran.
M.M.M.\ Fahim wrote the original draft; M.N.\ Molla, M.J.H.\ Imran,
M.M.R.\ Likhon, M.S.S.\ Saad, and M.R.\ Karim reviewed and edited the
manuscript. M.R.\ Karim supervised the project and provided resources and
project administration. All authors approved the final manuscript.


\end{document}